\documentclass{ws-procs9x6}

\begin{document}

\title{Dielectric function of the Si(113)3$\times$2ADI surface from
ab-initio methods}

\author{Katalin Ga\'al-Nagy$^*$ and Giovanni Onida}

\address{Dipartimento di Fisica and ETSF, Universit{\`a} degli Studi
  di Milano, via Celoria 16, I-20133 Milano, Italy
$^*$E-mail: katalin.gaal-nagy@physik.uni-regensburg.de\\
}

\begin{abstract}
We have investigated the imaginary part of the dielectric function
$\rm{Im}(\epsilon)$ of the (113) 3$\times$2 ADI reconstructed surface
of silicon. The calculations have been performed for a periodic slab
within the plane-wave pseudopotential approach to the
density-functional theory. The three diagonal components of
$\rm{Im}(\epsilon)$ have been derived from the momentum matrix
elements within the independent particle random phase approximation
(IPRPA). In this article, the importance of the {\bf k}-point
convergence is figured out by inspecting {\bf k}-point resolved
spectra.
\end{abstract}


\bodymatter


\section{Introduction}\label{Intro} 
The Si(113) surface is one of the most stable high-index surfaces of
Si \cite{Eag1993b}. It is used as a substrate for the self-assembled
growth of Ge nanowires and Ge as well as SiGe
islands\cite{Omi1999,Hal2002,Zha2004c,Han2005}. Since this surface is
atomically smooth, ultrathin oxide films can be grown on
Si(113)\cite{Mus2001a,Mus2001b}, which is hence also dealt as a
candidate for wafers and devices.

\begin{figure}[t]
  \begin{center}
  \psfig{file=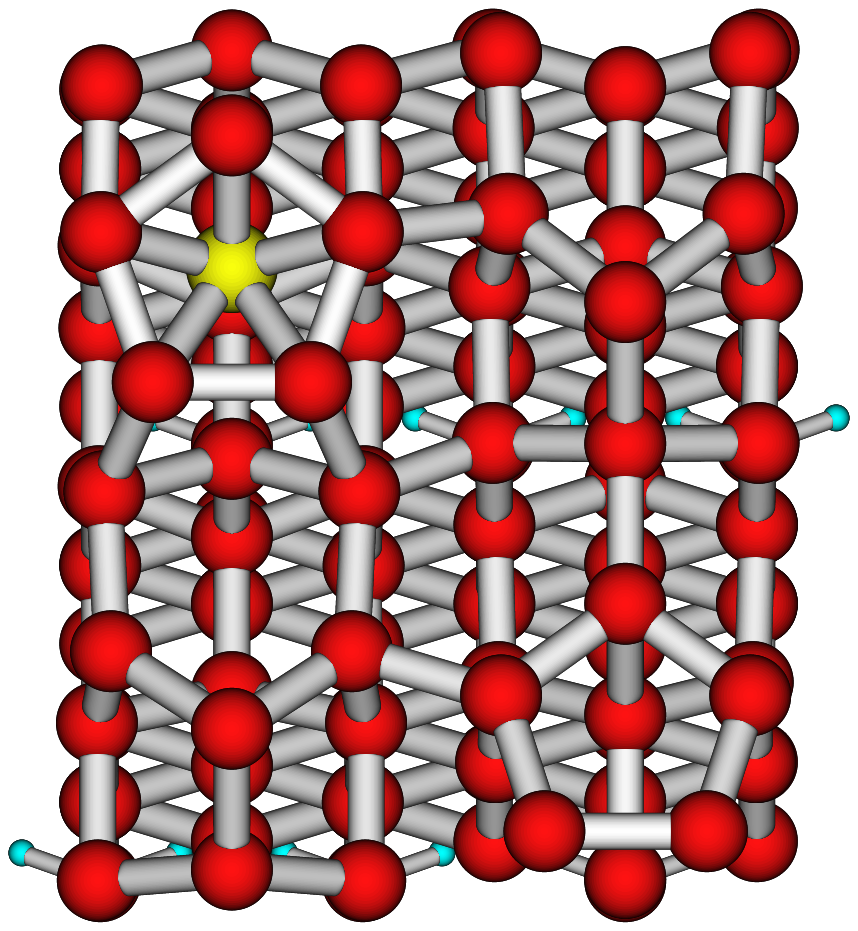,width=1.8in}
  \psfig{file=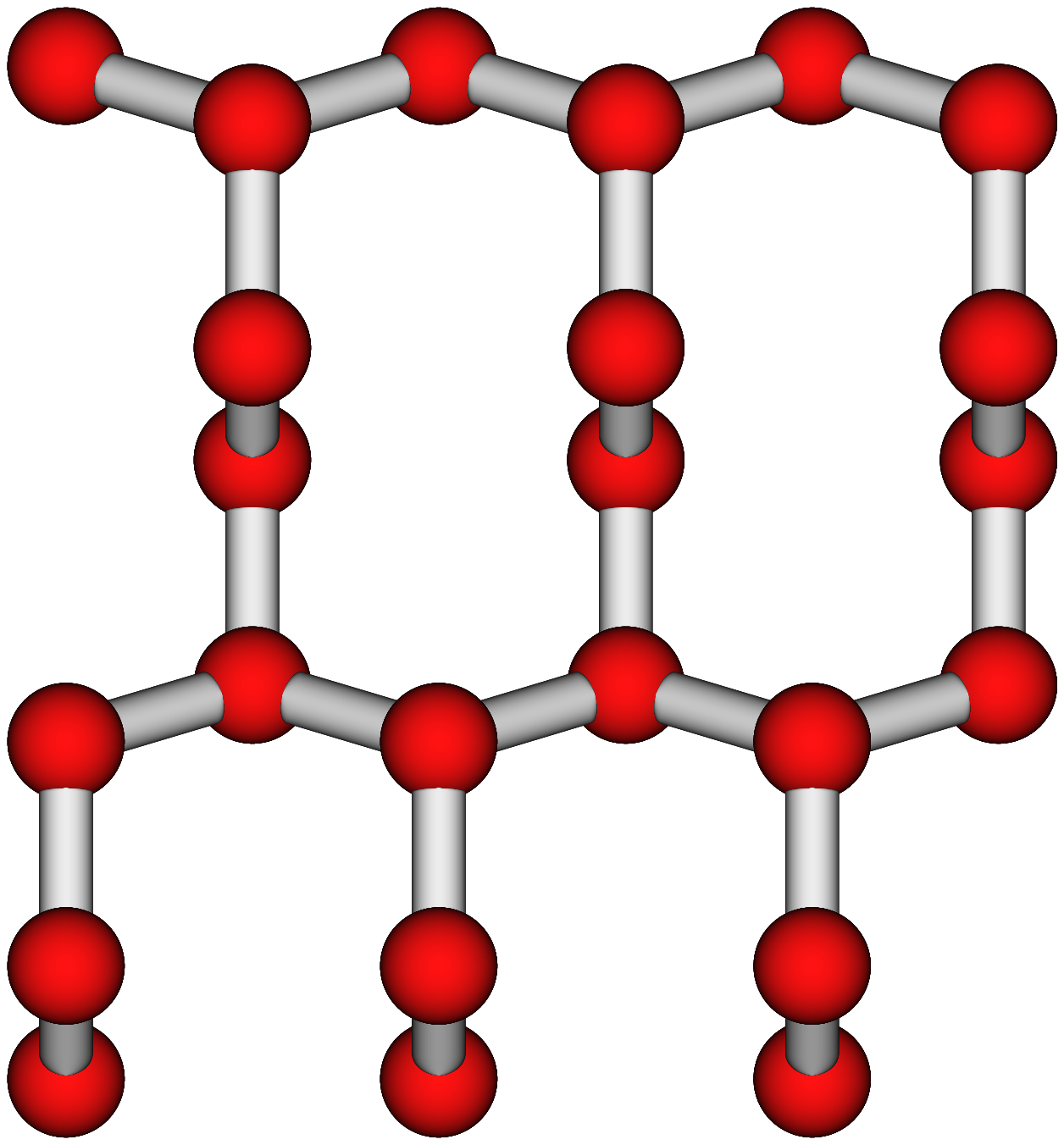,width=1.8in}
  \end{center}
  \caption{
  In the left panel, a top view ($x$-$y$ plane) of the
  Si(113)$3\times$2ADI reconstruction is shown. The right panel
  displays a top view of just two ``double layers'' of the bulk-like
  part of the slab.  The Si atoms are dark (the interstitial Si atom
  light) and the hydrogen atoms light.
  }\label{FigStruct}
\end{figure}

The clean surface shows a 3$\times$2 periodicity at room temperature
which can transform into a 3$\times$1 one by a phase transition
induced by high temperatures or
contamination\cite{Hwa2001,Jac1993}. The most probable is the
Si(113)3$\times$2ADI (adatom-dimer-interstitial) reconstruction, which
was proposed by D\c{a}browski et al.\cite{Dab1994} (see
Fig.~\ref{FigStruct}). Comparing ab-initio results for reasonable
surface models, the ADI reconstruction has the lowest surface
energy\cite{Ste2003a}. Besides, an experimental confirmation is given
by STM measurements\cite{Zha2004c,Kna1991}. However, still other
models can not be ruled out\cite{Kim2003}.

A complementary study of the Si(113)3$\times$2 surface can be done by
the investigation of its optical properties, e.g., the reflectance
anisotropy spectra (RAS). At the aim of performing a theoretical
ab-initio investigation of the optical properties the first step is
the calculation of the imaginary part of the dielectric function
$\rm{Im}(\epsilon)$. Results for $\rm{Im}(\epsilon)$ are presented in
this work.

%
%
\section{Method}\label{method} 
The calculations have been performed within the plane-wave (PW)
pseudopotential approach to the density-functional
theory\cite{Hoh64,Koh65} as implemented in ABINIT\cite{ABINIT} and
TOSCA\cite{TOSCA}. For the exchange-correlation energy the
local-density approximation has been used\cite{Per81,Cep80}. The
ground state convergence required 4 {\bf k} Monkhorst-Pack\cite{Mon76}
points in the irreducible wedge of the Brillouin zone (IBZ) as well as
a kinetic-energy cutoff of 12~Ry ($\approx$ 24000 PW). The slab used
for the calculation contains 11 double layers (DL) of Si where the
bottom surface is saturated with hydrogen atoms (see
Figs.~\ref{FigStruct} and \ref{FigLbL}). The topmost 4 DL have been
relaxed, and the remaining have been kept to bulk positions.

The imaginary part of the dielectric function ${\rm
Im}[\epsilon(\omega)]$ as a function of the energy $\omega$ has been
derived by a summation of the matrixelements of the momentum
operator\cite{Sol95}. The sum has to be taken over the valence- and
the conduction states (here: 175 and 185, respectively), and weighted
{\bf k} points, which in the present case can be taken just in one
quarter of the Brillouin Zone, which is the IBZ of the system.  For
the layer-by-layer analysis modified matrixelements have been used
according to the prescriptions of reference \refcite{Hog03} (see also
references \refcite{Cas03} and \refcite{Mon03}). Here, the effects of
local fields, self energy, and excitons have been neglected.

%
%
\section{{\bf k} point convergence}\label{convk}  

\begin{figure}[t]
  \begin{center}
  \psfig{file=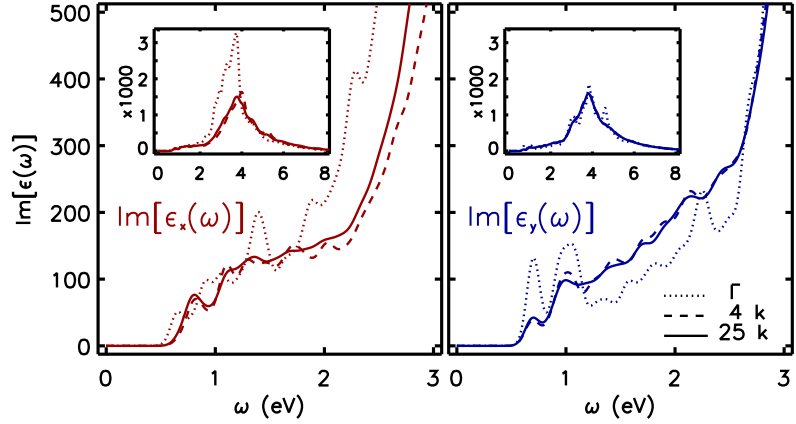,width=4.1in}
  \end{center}
  \caption{
  ${\rm Im}[\epsilon_x(\omega)]$ (left) and ${\rm
  Im}[\epsilon_y(\omega)]$ (right) for various numbers of {\bf k}
  points (assignment in the graph) for the low-energy range, where the
  surface states are found. The full energy range is shown in the
  insets.
  }\label{FigConvK}
\end{figure}

For the calculation of ${\rm Im}[\epsilon(\omega)]$ (and consequently,
for the calculation of all optical properties which are based on this)
the convergence with respect to number of {\bf k} points in the
summation described above is essential, since the spectra are
particularly sensitive to that. Working out of convergence can yield
wrong difference spectra (e.g., RAS). In comparison, the convergence
with respect to the number of bands is less demanding in the
surface-related low-energy range.  Thus, we performed a very careful
check of the {\bf k}-point convergence.

Since the slab was chosen with the surface perpendicular to $z$,
${\rm Im}[\epsilon(\omega)]$ should be investigated for light
polarized in $x$ and $y$ direction, since these functions are the
ingredients for difference spectra like the RAS. We show in
Fig.~\ref{FigConvK} the ${\rm Im}[\epsilon_x(\omega)]$ and ${\rm
Im}[\epsilon_y(\omega)]$ for various sets of {\bf k} points. For both
polarizations, at least 25 {\bf k}-points are required. The
convergence for ${\rm Im}[\epsilon_y(\omega)]$ is faster than for
${\rm Im}[\epsilon_x(\omega)]$. In particular, the differences between
4 and 25 {\bf k} points for the ${\rm Im}[\epsilon_x(\omega)]$ is
still significant.

\begin{figure}[t]
\begin{center}
  \psfig{file=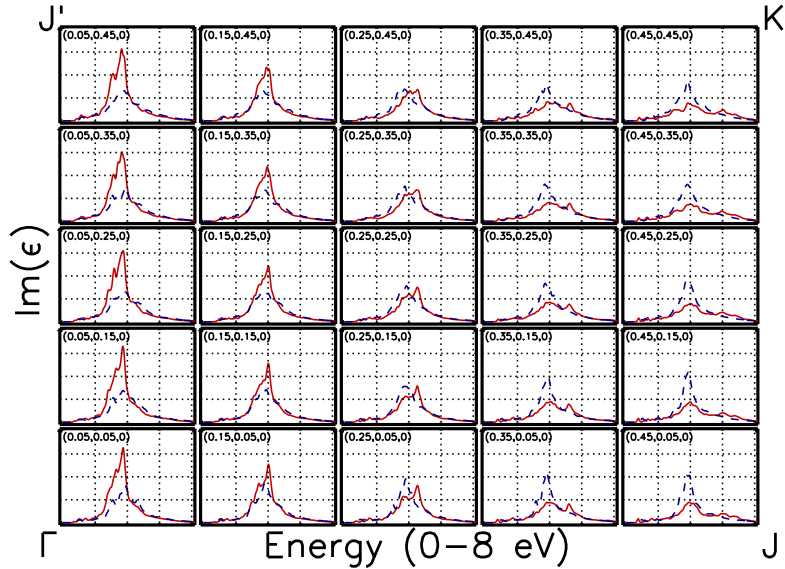,width=4.1in}
\end{center}
  \caption{
    Single-{\bf k} decomposition of the spectra using
    25 {\bf k} in the surface IBZ. ${\rm Im}[\epsilon_x(\omega)]$ is
    drawn with solid and ${\rm Im}[\epsilon_y(\omega)]$ with
    dashed lines. The coordinates of the {\bf k} points are
    denoted in the inset. The spectra are organized as the {\bf k}
    points in the surface IBZ, where the $\Gamma$ point is close to
    the low-left corner of the figure.
  }\label{FigSinglek}
\end{figure}

In order to improve the {\bf k}-points grid in the most efficient way,
we have performed a single-{\bf k} analysis. This means, we have
inspected the contributions of each {\bf k} point to the spectra
separately, which are shown in Fig.~\ref{FigSinglek}. It is apparent
that the changes for the ${\rm Im}[\epsilon_x(\omega)]$ from one {\bf
k} to a neighbouring one, especially in $x$ direction, are larger than
for ${\rm Im}[\epsilon_y(\omega)]$. Thus, we have chosen 30 additional
{\bf k} points in between along the $x$ direction. The spectra based
on the resulting 55 {\bf k} points (not shown in Fig.~\ref{FigConvK})
demonstrates, that convergence already had been achieved with 25 {\bf
k} points. This strategy allows one to avoid a ``blind'' increase of
the {\bf k}-point grid, which can be very time-consuming.

%
%
\section{Anisotropy at $\Gamma$}\label{gamma} 
\begin{figure}[t]
  \psfig{file=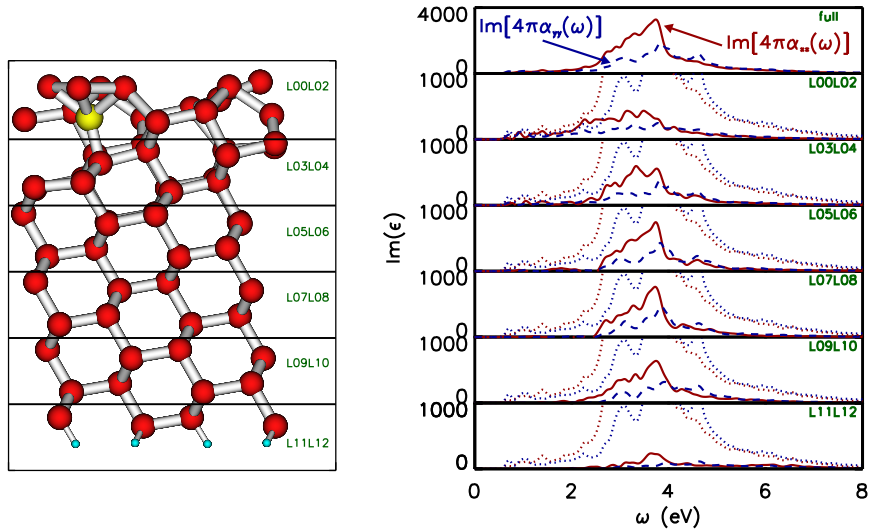,width=4.5in}
  \caption{
  Layer-by-layer decomposition of the spectra and
  complete spectra containing all layers (full) for the $\Gamma$ point
  calculation (right). The spectra correspond to the layers denoted in
  the inset and marked at the structure (left). ${\rm
  Im}[\epsilon_x(\omega)]$ is drawn with solid and ${\rm
  Im}[\epsilon_y(\omega)]$ with dashed lines. In addition, the
  full-slab spectra are drawn with dotted lines. DL L00 corresponds to
  the vacuum at the top of the slab and DL12 to the hydrogen at the
  bottom.
  }\label{FigLbL}
\end{figure}

In Fig.~\ref{FigConvK} a particular feature is visible: in the
contribution coming from the $\Gamma$ point, a strong anisotropy at
the medium-high energy range, between the $x$ and the $y$
polarization, occurs. This anisotropy at $\Gamma$ appears in the
energy range of bulk-bulk transitions. In fact, performing a
layer-by-layer analysis the spatial localization of such anisotropic
contributions is clearly found to be in the bulk (internal) DLs, as
shown in Fig.~\ref{FigLbL}, together with a side view of the slab cell
where the DLs used for each spectra are marked.  We have chosen always
pairs of DLs, e.g., L07L08 corresponds to the spectra with
contributions from DL~7 and DL~8, where the DLs are counted from the
top to the bottom.  Of course, bulk silicon is isotropic, which
results also from calculations using a sufficient amount of {\bf k}
points. The bulk anisotropy at the $\Gamma$ point here is just due to
the use of a low-symmetry supercell.

Fig.~\ref{FigLbL} shows that this anisotropy is present at each pair
of DLs in the bulk region of the slab, where the spectra for bulk
region, i.e. L09L10, L07L08, and L05L06, are nearly the same. The
explanation can be found by inspecting the geometry of one pair of DLs
only, which is shown on the right in Fig.~\ref{FigStruct}. An highly
anisotropic ``chain structure'' oriented along the $x$ axis can be
seen. This is the reason for the strong signal appearing at $\Gamma$
(the wavefunctions are in phase and the wavelength of light is much
larger than the cell dimension).

%
%
\section{Summary}\label{Summary} 
In summary, we have presented ab-initio results for the imaginary part
of the dielectric function for the Si(113)3$\times$2ADI surface. At
this example we have shown the possibility of accelerating the
convergence tests with respect to {\bf k} points by performing a
single-{\bf k} analysis. We have applied a layer-by-layer analysis
allowing us to explain a strong anisotropic contribution coming from
the $\Gamma$ point.  The results described here are useful for further
calculations, e.g., the RAS spectra, which can be obtained from ${\rm
Im}[\epsilon(\omega)]$.
%
%
\section{Acknowledgment}\label{dank}
This work was funded by the EU's 6th Framework Programme through the
NANOQUANTA Network of Excellence (NMP4-CT-2004-500198).  K.G.-N. also
likes to acknowledge S.~Hinrich and A.~Stekolnikov for support.

\bibliographystyle{ws-procs9x6}
\bibliography{Si113,Theory,Opt}

\end{document}